\documentstyle[12pt,aasms4,psfig,epsf]{article}

\received{April}
\accepted{July}
\revised{}
\singlespace

\begin{document}

\title{MAGNETIC FIELD EFFECTS ON THE STRUCTURE AND EVOLUTION
OF OVERDENSE RADIATIVELY COOLING  JETS}

\author{Adriano H. Cerqueira and Elisabete M. de Gouveia Dal Pino}

\affil{Instituto Astron\^{o}mico e Geof\'{\i}sico, Universidade de
S\~{a}o Paulo, \\ Av. Miguel St\'efano 4200, (04301-904) S\~{a}o Paulo
- SP, BRAZIL \\
adriano@iagusp.usp.br, dalpino@iagusp.usp.br \\}\vskip 2.0 cm

\centerline{\bf To appear in The Astrophysical Journal}

\begin{abstract}

We investigate the effect of magnetic fields on the propagation dynamics
and morphology of overdense, radiatively cooling, supermagnetosonic
jets, with the help of fully three-dimensional smooth particle
magnetohydrodynamic simulations. Evaluated for a set of parameters
which are mainly suitable for  protostellar jets (with density ratios
between the jet and the ambient medium $\eta \approx 3-10$, and ambient
Mach number $M_a \approx 24$), these simulations  are also compared
with baseline non-magnetic and adiabatic calculations.  Two initial
magnetic field topologies (in $\sim$ equipartition with the gas,
$\beta=p_{th}/p_B \simeq 1$) are considered: (i) a helical field and
(ii) a longitudinal field, both of which permeate both the jet and the
ambient medium.

We find that, after amplification by compression and re-orientation in
nonparallel shocks at the working surface, the magnetic field that is
carried backward with the shocked gas into the cocoon improves the jet
collimation relative to the purely hydrodynamic (HD) systems, but this
effect is larger in the presence of the helical field.  In both
magnetic configurations, low-amplitude, approximately equally spaced
($\lambda \approx 2-4R_j$) internal shocks (which are absent in the HD
systems) are produced  by MHD Kelvin-Helmholtz reflection pinch modes.
The longitudinal field  geometry also excites non-axisymmetric helical
modes which cause some beam wiggling.  The strength and amount of these
modes are, however, reduced (by $\sim$ twice) in  the presence of
radiative cooling relative to the adiabatic cases.  Besides, a large
density ratio, $\eta$, between the jet and the ambient medium also
reduces, in general, the number of the internal shocks.  As a
consequence, the weakness of the induced internal shocks makes it
doubtful that the magnetic pinches could produce by themselves the
bright knots observed in the overdense, radiatively cooling
protostellar jets.

Magnetic fields may leave also important signatures on the head
morphology of the radiative cooling jets.  The amplification of the
nonparallel components of the magnetic fields, particularly in the
helical field geometry, reduces the postshock compressibility and
increases the postshock cooling length. This may lead to stabilization
of the cold shell of shocked material that  develops at the head
against both the Rayleigh-Taylor and global thermal instabilities. As a
consequence, the clumps that develop by fragmentation of the shell in
the HD jets tend to be depleted in the helical field geometry. The jet
immersed in the longitudinal field, on the other hand, still retains
the clumps although they have their densities decreased relative to the
HD counterparts.  As stressed before (Cerqueira, de Gouveia Dal Pino \&
Herant 1997), since the fragmented shell structure resembles the knotty
pattern commonly observed in HH objects behind the bow shocks of
protostellar jets, this result suggests that, as long as
(equipartition) magnetic fields are present, they should  probably  be
predominantly longitudinal  at the  head of these jets.


\end{abstract}

\clearpage

\section{Introduction}

There is increasing evidence that protostellar jets are driven by
circumstellar magnetized disks associated with pre-main sequence stars
(e.g. K\"onigl \& Ruden 1993; Shu et al. 1995).  Efficient mass loss in
supersonic, collimated magnetized outflows is the most likely mechanism
by which protostars dissipate the angular momentum accumulated during
the accretion of the surrounding material. While of fundamental
importance in the production and initial collimation of the jets,
magnetic fields have been generally neglected in most of the analytical
and numerical modeling of the structure of protostellar jets since the
inferred estimates of their strength ($B \sim 10^{-6}~-~10^{-5}$ G)
suggested  that they could be not dynamically important along the flow
(e.g., Morse et al. 1993). However, recent observations of circularly
polarized radio emission of the young stellar outflow of T Tauri S (Ray
et al. 1997), for example, indicate the presence of a strong, ordered
magnetic field in the flow, far away from the source, which has
possibly been amplified by compression behind the shocks at the head of
the outflow.  After amplification and re-orientation behind the shocks,
such magnetic fields may operate significant changes on the  dynamics
and collimation of a jet and also in its  head structure as suggested
by recent numerical studies (see below).

Great effort has been concentrated in the analytical and numerical
study of  magnetized, adiabatic, light jets, aiming mostly the
investigation of extragalactic jets (see, e.g. Birkinshaw  1997 for a
review). Most of that work has focused  on the study of the stability
properties of the beam against hydromagnetic and Kelvin-Helmholtz
instabilities. In the limit of zero-velocity difference between the jet
and the surrounding medium, linear theory predicts that a jet
magnetically confined by a toroidal field is unstable to the pinch and
kink (or helical) {\it hydromagnetic} modes (e.g., Chandrasekhar 1961)
with the temporal growth rates of the pinching mode increasing with
increasing density ratio, $\eta$, between the jet and the ambient
medium. In the presence of a non-zero velocity discontinuity at the
boundary layer separating the two fluids, these pure hydromagnetic
modes are modified by the development of the Kelvin-Helmholtz (K-H)
instability. A jet confined by a toroidal magnetic field is unstable to
the fundamental and reflection pinch and kink modes of the K-H
instability (Cohn 1983; Fiedler \& Jones 1984).  The most unstable
pinching mode in this case, at wavelengths $\lambda \sim 2 \times \pi
R_j$ (where $R_j$ is the jet radius), has a destabilization length $l
\propto (M_j R_j)$ (where $M_j = v_j/c_j$ is the jet Mach number),
which is similar to the pure hydrodynamical (HD) case for jets with
$\eta ~{\rm <}~ (M_j/2)^{3.3}$ (Cohn 1983). In the case of magnetized
jets with a longitudinal magnetic field residing in an unmagnetized
medium, all K-H modes become stable for sub-Alfv\'enic flows (except
for a small region of slow reflection modes; e.g., Bodo et al. 1989;
Hardee et al. 1992; Hardee, Clarke \& Rosen 1997). In the super-Alfv\'enic
regime, on the other hand, the growth rates of the instability are not
very much different from those of pure HD flows.  For example, a
fundamental kink mode in the pure HD case can be identified with an
Alfv\'en disturbance of long wavelength in the magnetohydrodynamic
(MHD) case, and reflection modes at shorter wavelengths can be
identified with fast magnetosonic (ms) waves reflecting off the jet
boundaries. In this super-Alfv\'enic regime, if the jet is also
supermagnetosonic ($M_{ms,j} = v_j/(v_A^2 + c_j^2)^{1/2} > 1$, where
$v_A$ is the Alfv\'en velocity and $c_j$ is the jet sound speed), then
it becomes more stable with increasing $M_{ms,j}$, with the
destabilization length varying approximately proportional to $M_{ms,j}$
($l \propto M_{ms,j} R_j$) (Hardee et al.  1992).  Also, strong
toroidal fields of strength comparable to the poloidal field can lead
to increased jet stability (Appl \& Camenzind 1992; Thiele \& Camenzind
1997).

The effects of the K-H modes on the survival of the beam, however,
cannot be predicted by the linear theory alone. As in pure HD flows,
the fastest growing, shortest wavelength reflection modes are expected
to saturate with the formation of weak oblique shocks, while the
fundamental modes may not saturate and may cause large-scale
distortions and even disruption of the flow. Numerical simulations
which can determine the end points of the operation of these
instabilities confirm these predictions. Hardee et al. (1992) and
Hardee, Clarke \& Rosen (1997), for example, assuming slab and
cylindrical jets, respectively, with axial magnetic fields, have
focused on the comparison of the scale-length of the structures
generated during the nonlinear evolution and the wavelengths of maximum
growth rate predicted by the linear theory. They find that the jet is
not stabilized by nonlinear effects associated with increasing magnetic
tension and disrupts near the $resonant$ wavelength of the K-H kink
mode. Malagoli, Bodo \& Rosner (1996) and Min (1997a, b) have extended those
investigations of the nonlinear development of the K-H instability by
including diffusion and magnetic reconnection effects and found that
even if the magnetic field intensity is not too large to completely
suppress the K-H instability, it is still able to mediate turbulence
decay and diffusion of energy and mass across the boundary layer
between jet and surrounding medium.

These studies of the stability properties of the beam against MHD  K-H
pinch and helical modes seem to provide  potential mechanisms to
explain the formation of structures such as knots and wiggles in
adiabatic, supermagnetosonic light jets. Besides these stability
analyses, some effort has  also been spent in numerical studies of the
general effects of ${\bf B}$-fields on the global evolution and
morphology of adiabatic, light  jets, still in the context of
extragalactic jets, assuming both passive (Clarke, Norman \& Burns
1989; Matthews \& Scheuer 1990; Hardee \& Norman 1990) and dynamical
important toroidal fields (e.g., Clarke , Norman \& Burns 1986; Lind et
al. 1989; K\"ossl, M\"uller \& Hillebrandt 1990). In the later case,
the jet was found to be rapidly decelerated at the Mach disk with the
shocked jet material being confined to a slender trans-Alfv\'enic plug
instead of being deposited in the large cocoon observed in the purely
HD light jets.

Lately, these numerical MHD studies have been extended to heavy,
adiabatic jets (Todo et al. 1993; Hardee \& Clarke 1995; Stone, Xu \&
Hardee 1997). Hardee \& Clarke and Stone et al. have focused on
simulations of jets with a poloidal field propagating into an
unmagnetized medium, while Todo et al. (see also Thiele \& Camenzind
1997) have considered a helical field configuration extending also to
the ambient medium.  Comparing the development of supermagnetosonic,
adiabatic, heavy jets with axial and toroidal magnetic fields
propagating into unmagnetized ambient media, Hardee \& Stone (1997)
have found that the toroidal geometry suppresses the mixing and
entrainment of ambient gas which is found to develop in the axial case
as a consequence of the K-H instability. Also, they have found that the
kink mode has a longer wavelength and smaller amplitude in the toroidal
configuration.

Still almost unexplored, however, is the role played by ${\bf
B}$-fields on the propagation dynamics and  morphology of radiatively
cooling, heavy  jets  (a scenario which is believed to occur in
protostellar jets).  In the limit of zero magnetic field, numerical
simulations of  radiatively cooling, heavy jets [e.g., Blondin, Fryxell
and K\"onigl 1990, hereafter BFK; de Gouveia Dal Pino \& Benz 1993,
hereafter GB93; Stone \& Norman 1993a, 1993b, 1994; de Gouveia Dal Pino
\& Benz 1994, hereafter GB94; Chernin et al. 1994; de Gouveia Dal Pino,
Birkinshaw \& Benz 1996, hereafter GBB96; de Gouveia Dal Pino \&
Birkinshaw 1996 (GB96)] have shown that thermal energy losses by the jet
system have important effects on its dynamics.  These studies have
revealed that a cooling jet develops a dense, cold shell of shocked
material at the head which is fragmented into clumps by Rayleigh-Taylor
instability.  Moreover, BFK and GB93 have found that the development of
the K-H modes along the beam are inhibited by the presence of cooling.
Recently, Hardee \& Stone (1997) and Stone, Xu \& Hardee (1997) (see
also Massaglia et al. 1992) have examined the dynamics of K-H unstable
cooling jets. Their linear analysis indicate that the growth of the K-H
modes is very sensitive to the assumed form of the cooling function.
In particular, if the cooling curve is a steep function of the
temperature in the neighborhood of the equilibrium state, then the
growth of K-H modes is reduced relative to the adiabatic jet - a result
which is consistent with previous numerical simulations (BFK; GB93).
With the inclusion of a longitudinal magnetic field in a
supermagnetosonic jet, Hardee \& Stone (1997) find that the magnetic
field does not strongly impact the differences in the K-H stability
properties  between an adiabatic and a cooling HD jet, provided that
the magnetic pressure does not dominate.  Besides, they find that the
increase in the magnetic field strength makes the linear stability
properties to become more like those of an adiabatic jet.  The
nonlinear analysis of the growth of the K-H modes in cooling jets in
the $B=0$ limit shows similar behaviour as in the pure adiabatic cases.
The jet can disrupt near the resonant frequency of the fundamental kink
K-H mode, and nonlinear, higher frequency, reflection waves tend to
produce low-amplitude wiggles, and can result in strong shocks in the
jet beam.

In the present work, we attempt to extend these previous investigations
by exploring the nonlinear effects of magnetic fields (close to
equipartition with the gas) on the global evolution and morphology of
radiatively cooling, heavy jets and test the morphological signatures
of two different magnetic field geometries (a longitudinal and a
helical configuration) on the  dynamics of protostellar jets.  A
preliminary step in this direction was made in  previous work (de
Gouveia Dal Pino \& Cerqueira 1996; Cerqueira, de Gouveia Dal Pino \&
Herant 1997, hereafter Paper I), where we have mainly focused on the
effects of the magnetic fields in the jet $head$ structure.  With the
help of three-dimensional (3-D) smooth particle magnetohydrodynamical
(SPMHD) simulations, we have found that the presence of a helical
magnetic field (in close equipartition with the gas) may suppress the
formation of the clumpy structure that is found to develop at the head
of HD jets by fragmentation of the cold shell of shocked jet material.
A cooling jet immersed in a longitudinal magnetic field, on the other
hand, tends to retain the clumpy morphology at its head.  In the
present work, we perform 3-D SPMHD simulations to address the details
of the magnetic field effects over the whole structure of the
radiatively cooling, heavy jets, covering a more extensive range of
parameters, and compare with both non-magnetic and adiabatic systems.

Contemporaneously with this work, Frank et al. (1998; see also Frank et
al.  1997) have performed grid-based 2$\frac{1}{2}$-D simulations of
magnetized, radiatively cooling jets assuming a toroidal ($B_{\phi}$)
magnetic field and concluded that toroidal fields are able to excite
the development of strong pinches  along the beam. Although in our
simulations we have assumed somewhat different initial conditions (see
below), where both investigations overlap, the   results are
qualitatively similar, except for the fact that in our simulations the
MHD pinch modes are found to be not strong in radiatively cooling,
heavy jets.  This difference, however,  is mainly due to differences in
the assumed initial density ratios between the jet and the ambient
medium, which are much larger in our analyses (see Paper I and
discussion below).

In \S 2, the numerical method and the initial conditions used are
briefly described. In \S 3, we present the results of our simulations,
and in \S 4, we address the conclusions and the possible implications
of our results.

\section{Numerical Technique}

In our simulations, we consider the magnetohydrodynamics (MHD)
conservation  equations in the ideal approximation

$$\frac{d{\rho}}{dt}=-\rho{\bf {\nabla}}\cdot{\bf v} \eqno(1a)$$

$$\frac{d{\bf v}}{dt}=-\frac{{\bf
{\nabla}}p}{\rho}+\frac{1}{4{\pi}{\rho}}( {\bf {\nabla}}\times{\bf
B})\times{\bf B} \eqno(1b)$$

$$\frac{du}{dt}=-\frac{p}{\rho}({\bf {\nabla}}\cdot{\bf v}) - {\cal L}
\eqno(1c)$$

$$\frac{d{\bf B}}{dt}=-{\bf B}({\bf {\nabla}}\cdot{\bf v})+
({\bf B}\cdot{\bf {\nabla}}){\bf v} \eqno(1d)$$

\noindent where the symbols have their usual meaning (i.e., $\rho$ is
the density; ${\bf B}$ is the magnetic field; $u$ is the specific
internal energy, ${\cal L}$ is the radiative cooling rate, etc.).  To
close the system (1), an ideal equation of state is assumed:

$$p=(\gamma -1)\rho u \eqno(1e)$$

\noindent with $\gamma=5/3$.

The MHD equations above are solved (in Cartesian coordinates) using a
modified version of the fully three-dimensional smooth particle
hydrodynamics (SPH) code originally developed by de Gouveia Dal Pino \&
Benz (1993) (see also GB94; Chernin et al.  1994; GB96; GBB96) for the
investigation of the evolution of purely hydrodynamic (HD) jets.

In the SPH formalism, equations (1) are described by (e.g., Benz 1990;
Stellingwerf \& Peterkin 1990; Monaghan 1992; Meglicki 1995):

$$
\bigg\langle\frac{d{\bf v}}{dt}\bigg\rangle=-\sum_{j=1}^N m_j \bigg(
\frac{{p_i}}{{\rho}_i^2} +\frac{{p_j}}{{\rho}_j^2}+\Pi_{ij}\bigg) {\bf
{\nabla}}_iW_{ij}+
$$
$$
~~~~~~~~~~~~~~~~~~~~~~~~~~~~~~~~~~~~~~+\frac{1}{{4{\pi}{{\rho}_i^2}}}
\sum_{j=1}^N \lbrack m_j ({\bf B}_i - {\bf B}_j)\times {\bf {\nabla}}_i
W_{ij} \rbrack \times {\bf B}_i \eqno(2a)$$

$$\bigg\langle\frac{du}{dt}\bigg\rangle=\frac{p_i}{{\rho}_i^2}
\sum_{j=1}^N m_j({\bf v}_i-{\bf v}_j){\bf {\nabla}}_i W_{ij}+
\frac{1}{2}\sum_{j=1}^N m_j \Pi_{ij}({\bf v}_i-{\bf v}_j)\cdot {\bf
{\nabla}}_i W_{ij} \eqno(2b)$$

$$ \bigg\langle\frac{dB_k}{dt}\bigg\rangle=\frac{1}{\rho_i}
\sum_{j=1}^N m_j (B_{i{\rm ,}k}{\bf v}_{ij}-v_{ij{\rm ,}k}{\bf
B}_i){\bf {\nabla}}_iW_{ij} \eqno(2c)$$

\noindent where $m_j$ is the mass of the particle $j$ (located at the
position ${\bf r}= {\bf r}_j$); $\rho_i$ and $\rho_j$ are the density
of the particles $i$ and $j$, respectively; $p_i$ and $p_j$ are their
pressure; ${\bf B}_i$ is the magnetic field at ${\bf r}_i$ ($B_{i,k}$
is the $k$ component of ${\bf B}_i$); ${\bf v}_{ij}\equiv {\bf
v}_i-{\bf v}_j$ is the velocity difference, and $v_{ij,k}$ is $k$
component of ${\bf v}_{ij}$.  The brackets on the left-hand side of the
equations mean that the physical quantities inside them are evaluated
at the position $ {\bf r}_i$ (at the particle $i$).  The term ${\bf
{\nabla}}_iW_{ij}$ is the gradient of the Kernel function $W_{ij}$ at
the position of the particle $i$, and $\Pi_{ij}$ is the artificial
viscosity which allows for appropriate treatment of  shock wave
dissipation.  We here adopt the von Neumann-Richtmyer viscosity (e.g.,
Benz 1990).

The density and pressure are both calculated from the definition of
discretness in SPH (e.g., Benz 1990; Monaghan 1992), and are
expressed, respectively, by the equations:

$$\langle \rho \rangle= \sum_{j=1}^N m_jW_{ij} \eqno(2d)$$

$$\langle p \rangle= (\gamma -1) \sum_{j=1}^N m_j u_j W_{ij}
\eqno(2e)$$

As in previous work (GB93; GB94; GB96), the radiative cooling rate (due
to collisional excitation and recombination) ${\cal L}$ in equation
(1c), is implicitly calculated using the  cooling function evaluated
by  Katz (1989)  for a gas of cosmic abundances cooling from from $T
\simeq 10^6$ K to $T \simeq 10^4$ K. [The cooling is suppressed below
$T \approx 10^4$ K, where the transfer of ionizing radiation becomes
important, and the assumption of a fully ionized flow breaks down (see,
e.g., GB93; GBB96)].

The choice of a {\it vectorial} formalism to write the system of
equations above (instead of a tensorial form) is to ensure that the
magnetic force vector [($\nabla \times {\bf B}$)$\times{\bf B}$], in
each point of the system, is strictly perpendicular to the magnetic
field vector itself.  Tensorial implementations in SPMHD, like those
proposed by Monaghan (1992), are subject to the development of
non-physical components of the magnetic force, whose projection on the
magnetic field vector has non-zero values. Furthermore, these parallel
components of the magnetic force are proportional to $\nabla \cdot {\bf
B}$ (see Brackbill \& Barnes 1980).  To avoid this, we have used the
vectorial form and found that those non-physical accelerations are
absent in our runs [i.e.,  ${\bf F}\cdot {\bf B} \simeq
10^{-6}-10^{-7}$, where ${\bf F}=\frac{1}{4\pi \rho} (\nabla \times
{\bf B})\times {\bf B}$].

There are several methods which can work well in order to maintain the
divergence of the magnetic field close to the  machine roundoff error
in grid-based codes (e.g., Evans \& Hawley 1988; Schmidt-Voigt 1989).
However, the maintenance of $\nabla \cdot {\bf B} = 0$ in SPMHD is
not an easy task. A {\it divergence-cleaning} process (e.g.,
Schmidt-Voigt 1989; Otmianowska-Mazur \& Chiba 1995), for example, is
not applicable yet in the context of the SPMHD, and the best way we
have found to diminish the effects of a potential
$\nabla \cdot {\bf B} \ne 0$ was to avoid those non-physical
accelerations by writing our system of equations in the vector
formalism.  Furthermore, we track the behaviour of
$\nabla \cdot {\bf B}$ by evaluating the following quantity  at
each time step and position of the system (e.g, Otmianowska-Mazur \&
Chiba 1995)

$$ \omega = \frac{\vert \nabla \cdot {\bf B}\vert \cdot h}{\vert
{\bf B} \vert}$$

As in Otmianowska-Mazur \& Chiba 1995 (see also Meglicki 1994, 1995 for
a detailed discussion), we have set $\omega \le 10^{-2}$ as a limit to
the validity of our calculations. In general, 85\% of the particles of
the system keep $\omega \le 10^{-2}$ (with almost all of them
distributed around the zero value most of the time). For small periods
of time, some of them may acquire an $\omega > 10^{-2}$, but the scalar
product between the magnetic field and the magnetic force remains very
small ($ \sim 10^{-6}-10^{-7}$, in code units) over the entire
evolution of the systems simulated here.

As additional tests to check the validity of the code modified by the
{\bf B}-field implementation, we have run  Alfv\'en wave tests, similar
to those suggested by Clarke (1996). In the centre of a rectangular
box, a circular pulse of radius  $r=\sqrt{y^2+z^2}=5h$  (where $h$ is
the smoothing length) and velocity ${\bf v}=v_{init}${\bf i} is set
perpendicularly to the magnetic field. The latter was assumed to be in
the $y$ direction in one case (${\bf B}=B_{init}${\bf j}) and ${\bf
B}=B_{init}\cdot$({\bf j} $+$ {\bf k}) in the other case. The initial
velocity was set to  $v_{init}/v_{A}=\sqrt{4\pi \rho}\cdot
v_{init}/B_{init}=10^{-3}$ and $v_{init}/v_{A}=(\sqrt{2}/2)\sqrt{4\pi
\rho}\cdot v_{init}/B_{init}=(\sqrt{2}/2)10^{-3}$, for each case,
respectively (see  Clarke 1996). Similarly to Clarke's tests, after
several time steps the diffusion of the pulses was  found to be
strictly confined to the direction of propagation of the Alfv\'en
waves, as depicted in Figs. 17 and 18 of Clarke (1996).

\subsection{Initial and boundary conditions}

The computational domain is represented by a 3-D rectangular box of
dimensions $-17R_j \le$ x $\le 17R_j$, $-22R_j \le$ y,z $\le 22R_j$,
where $R_j$ is the initial jet radius ($R_j$ is the code distance
unit). The Cartesian coordinate system has its origin at the  center of
the box and the jet  flows through the x-axis, and is continuously
injected into the bottom of the box [at ${\bf r}=(-17R_j,0,0)$]. Inside
the box, the particles are initially distributed on a cubic lattice. An
outflow boundary condition is assumed for the boundaries of the box.
The particles are smoothed out by a spherically symmetric kernel
function of width $h$, and the initial values of $h$ were chosen to be
$0.4 R_j$ and $0.2R_j$ for the ambient and jet particles,
respectively.

We consider two different initial magnetic field configurations.  One
of them is an initially constant  longitudinal ${\bf B}$-field permeating both
the jet and the ambient medium  [${\bf B}=(B_{x_0},0,0)$]. An
observational support for this kind of configuration is suggested by
the fact that some protostellar jets appear to be aligned with the
main direction of the local interstellar magnetic field (e.g.,
Appenzeller 1989).  The other adopted configuration is a force-free
helical magnetic field which also extends to the ambient medium with a
functional dependence  given by (see also Todo et al. 1993):

$$B_r=0 \eqno(3a)$$

$$B_{{\phi}}(r)=B_0{\bigg[} \frac{0.5Ad r^2}{(r+0.5d)^3}{\bigg]}^{1/2}
\eqno(3b)$$

$$B_x(r)=B_0{\bigg[}1- \frac{Ar^2(r + d )}{(r+0.5 d )^3} {\bigg]}^{1/2}
\eqno(3c)$$

\noindent where $r=\sqrt{y^2+z^2}$ is the radial distance from the jet
axis and the (arbitrary) constants $A$ and $d$ are given by 0.99 and
$3R_j$, respectively. In these equations, $B_0$ is the maximum strength
of the magnetic field and corresponds to the magnitude of the
longitudinal component at the jet axis. Figure 1 displays both the
longitudinal ($B_x$) and the toroidal ($B_{{\phi}}$) components as a
function of the radial distance (in code units, for $\beta=1$ at the
jet axis). The pitch angle at $1 R_j$ is $\approx 19^{\circ}$.

The models are parameterized by the dimensionless numbers:  {\it i})
the density ratio between the jet and the ambient medium,
$\eta=n_j/n_a$; {\it ii}) the ambient Mach number, $M_a=v_j/c_a$ (where
$v_j$ is the jet velocity and $c_a$ is the ambient sound speed); {\it
iii}) the jet to the ambient medium pressure ratio at the jet inlet,
$\kappa=p_j/p_a$, that we assume to be equal to unit; {\it iv}) the
thermal to the magnetic pressure ratio, at the jet axis
$\beta=p_{th}/p_b$; and {\it v}) $q_{bs} = d_{cool}/R_j$, the ratio of
the cooling length in the post-shocked gas behind the bow shock to the
jet radius (see, e.g., GB93).

\section{The Simulations}

As in previous work (e.g., GB96), based on typical conditions found in
protostellar jets, we have adopted the following initial values for the
parameters:  $\eta = 3 - 10$ (e.g., Morse et al. 1992; Raga \&
Noriega-Crespo 1993), an ambient number density $n_a = 200$ cm$^3$,
$v_j = 398$ km s$^{-1}$ (e.g., Reipurth, Raga \& Heathcote 1992), $M_a
= 24$, and $R_j = 2 \times 10^{15}$ cm (e.g., Raga 1993).  In the MHD
simulations, we have assumed an initial $\beta = 1$ (which corresponds
to a maximum initial value $B_0 = 83~\mu$ G). The subsections below
present the results of the simulations we have performed for both
radiatively cooling and adiabatic jets with and without magnetic
fields, and Table 1 summarizes the values of the input parameters. In
Table 1, $M_{A_j}=v_j/v_{A_j}$ and $M_{{ms}_j} =
v_j/\sqrt{v_{A_j}^2+c_j^2}$, give the initial Alfv\'en and magnetosonic
Mach numbers, respectively.  The purely hydrodynamical models are
labeled with  ``HD" and the MHD models with ``ML" and ``MH" in the
cases of the models with an initial longitudinal magnetic field and an
initial helical field, respectively.

\subsection{Radiatively cooling jets}

Figure 2 depicts  the time evolution of the magnetic field distribution
of two supermagnetosonic, radiatively cooling jets:  ML3r (an MHD model
with initial longitudinal magnetic field configuration; top) and MH3r
(an MHD model with initial helical magnetic field configuration;
bottom).  Both models have  initial $\eta=n_j/n_a=3$  and ambient Mach
number $M_a=24$ (see Table 1).  We find that the field lines are
amplified and reoriented across the bow shock in both magnetic field
configurations. The shocked jet material which is decelerated at the
head and is deposited in the cocoon carries the field lines embedded in
it. Part of these lines have their polarity reversed as we can see in
Fig. 2.

The cooling parameters behind the bow shock, $q_{bs} \simeq 8.0$, and
the jet shock, $q_{js} \simeq \eta^{-3}q_{bs} \simeq 0.3$ (see, e.g.,
GB93) in Fig. 2, imply that, within the head of the jet, the ambient
shocked gas is almost adiabatic, whereas the shocked jet material is
subject to rapid radiative cooling. The corresponding density contours
and velocity field distribution maps of the models above were presented
in Figs. 1 and 2 of Paper I, where they were also compared with a pure
hydrodynamical model (HD3r; see also Table 1). As we have stressed in
Paper I, the cold dense shell that develops at the jet head due to the
cooling of the shock-heated jet material in the pure hydrodynamical
case, also appears in the MHD jets.  Similarly, it becomes
Rayleigh-Taylor (R-T) unstable (e.g., GB93) and breaks into blobs that
spill into the cocoon and show a resemblance with the Herbig-Haro
objects which are observed at the head of protostellar jets.  As in the
HD case, the density in the shell of the MHD jets also undergoes
fluctuations with time which are caused by {\it global thermal
instabilities} of the radiative shock (see e.g., Gaetz, Edgar \&
Chevalier 1988 and GB93 and for details). Fig. 3 shows the time
evolution of the shell density ($n_{sh}$) variations at the jet axis of
the three jets, which have a period of the order of the gas cooling
time ($t_{cool} \approx 10$ years, or $t_{cool} \approx 0.3 t_d$, where
$t_d=R_j/c_a \approx 38$ years corresponds to the transverse jet
dynamical time).  We note that although the HD and the MHD jets attain
a maximum density of approximately the same magnitude at $t/t_d \sim
1.1$, later on the density enhancement in the MHD jets is inhibited by
the presence of the {\bf B}-field, particularly in the jet with helical
field (MH3r).

Fig. 2 indicates that the cold blobs that develop from shell
fragmentation at the head of the MHD jets detach from the beam as they
are expelled backward to the cocoon.  The survival of these blobs in
the cocoon seems to suffer with the presence of the magnetic fields,
particularly in the helical case.  Compared to the blobs of the HD jet,
their density is reduced by a factor $\sim 2$ in the longitudinal case
(ML3r) at the final time step ($t/t_d=1.65$), and almost vanishes in
the helical jet (MH3r).  Fig. 4 compares the density and pressure
profiles across the flow in three different positions along the MHD
jets in the head region. The lateral blobs that develop in the cocoon,
on both sides of the beam are clearly less intense in the helical case
than in the longitudinal case.  In the helical case, for example, we
find that the toroidal (nonparallel-to-shock) component of the magnetic
field, which is initially less intense than the longitudinal component
(by a factor $\le$ 2.5), is strongly amplified by compression in the
shocks at the head (by a factor $\sim 5$). As pointed out in Paper I,
this amplification increases  the cooling length behind the jet shock
and reduces the density growth. As a consequence, the shell tends to
stabilize against the R-T instability and the clump formation is
inhibited. At the end of the evolution, the dense shell with clumpy
structure which was observed to develop in the head region of the pure
HD jet (Fig. 1 of Paper I) is replaced in the helical case by an
elongated plug (Fig. 2) of low density material.

Fig. 2 also indicates the development of some pinching along both MHD
jets. In the pure HD case, constriction occurs only very close to the
jet head where the beam is over-confined by the gas pressure of the
cocoon (see Figs. 1 and 2 of Paper I; see also Fig. 5b below).  In the
MHD jet with helical field (MH3r), the toroidal component ($B_{\phi}$),
which is amplified by compression in the shocks at the head, is
advected back with the shocked material to the cocoon. The associated
magnetic pressure ($\sim B_{\phi}^2/8 \pi$) causes an increase in the
total pressure of the cocoon relative to the pure HD case, which
collimates the beam and excites the (fastest growing) small-wavelength
pinch modes of the MHD K-H instability. These modes over-confine the
beam and drive the approximately equally spaced internal shocks seen in
the MH3r jet (Fig.  2; bottom).  Along the MHD jet with longitudinal
field (ML3r; Fig. 2; top), the increase in the total confining pressure
of the cocoon also drives the development of the MHD K-H instabilities
which excite beam pinching and internal shocks.  Consistent with linear
theory for K-H modes, in the supermagnetosonic regime considered here,
they begin to appear at a distance $\sim M_{{ms}_j}R_j$, which is
smaller than in the pure HD jet (see Tabel 1).

The presence of these oblique internal shocks along the beam of the MHD
jets can be testfied by the density and pressure profiles across and
along the flow depicted in Fig. 5, where the pure HD  model is also
depicted for comparison. We see that the induced internal shocks in the
MHD cases have a density contrast $n_{is}/n_j \approx 4$ in the ML3r
model (middle) and $n_{is}/n_j \approx 5$ in the MH3r model (or
$n_{is}/n_{sh} \simeq 0.08$ and $0.15$, respectively, where $n_{sh}$ is
the density at the shell).  We also note a close correlation between
the pinching zones and the appearance of more intense reversed fields
in the contact discontinuity between the jet and the cocoon. In both
magnetic field configurations, the original longitudinal components are
reoriented in the nonparallel shocks at the head  and advected back to
the cocoon.  As a consequence, a predominantly toroidal current density
distribution (${\bf J}_{\phi}$) develops around the jet. Such
configuration creates a ${\bf J} \times {\bf B}$ force
($-J_{\phi}B_{\parallel}$) that constricts the beam triggering the MHD
pinch formation.

Fig. 6 depicts the density in the midplane section of the head of
three  supermagnetosonic, radiatively cooling jets with $\eta=10$ after
they have propagated over a distance $\sim 33R_j$: a purely HD model
HD10r (top), an MHD model with initial longitudinal magnetic field
configuration, ML10r (middle), and an MHD model with initial helical
magnetic field configuration MH10r (bottom). Figures 7 and 8 show the
time evolution of the corresponding velocity and magnetic field
distributions. The other initial conditions are the same as in Fig. 2.
For the time evolution depicted, there is no significant difference
between the MHD jets with different $\eta$ other than those presented
by their pure HD counterparts. In both HD and MHD cases, the jet with
larger $\eta$ [and thus with $v_{bs} \simeq v_j/(1-{\eta}^{-1/2})
\rightarrow v_j$] plows into the ambient medium almost ballistically
without accumulating much waste of shocked jet gas in the cocoon. The
shocked ambient material is almost adiabatic ($q_{bs}>>1$; see Table 1)
and accumulates in the shroud that envelops the cocoon/beam, while the
shocked jet material cools much faster ($q_{js}<< 1$) and the shock is
effectively isothermal.  The cold shell is thus much thinner in the
larger $\eta$ jets and the head resembles a bullet.

As in the $\eta=3$ case, the density in the shell of the $\eta=10$ jets
also undergoes variations with time (with a period of the order of the
jet radiative cooling time) which are caused by the global thermal
instabilities of the radiative shocks.  Likewise, after reaching a
similar maximum density amplitude, the MHD jets have their shell
density growth inhibited due to the decrease of the shock
compressibility caused by the presence of the {\bf B}-field,
particularly in the jet with helical field (MH10r).
We note, however, that in the HD jet, the shell density variations
attain a smaller maximum density amplitude than in the MHD cases. This
is possibly due to the smaller total pressure confinement that acts on
its head.

In the $\eta=10$ MHD jets (ML10r and MH10r), the increase in the total
confining pressure in the cocoon (due to the presence of the magnetic
field) also over-confines the beam, relative to the HD jet, and drives
some beam pinching.

Fig. 9a shows the density and pressure profiles across the flow in
different pinch positions along the MHD and HD $\eta = 10$ jets after
they have propagated over a distance $33R_j$ (at $t/t_d=1.65$), which
can be compared with Fig. 5 (for $\eta=3$).  As in the $\eta=3$ models,
the pinch collimation is larger in the helical MH10r jet than in the
ML10r jet, while the pure hydrodynamic jet, HD10r, has not developed
any internal pinches over the time scale depicted.  It is interesting
to note that in the MHD jets, the fastest-growing pinch modes of the
K-H instability would be expected to appear only at distances $\sim
M_{ms_j}R_j \sim 50R_j$, which are beyond the computed scales.  Thus
the early development of pinches in these cases, is possibly being
triggered mainly by the hydromagnetic $\theta$-pinch effect (e.g., Boyd
\& Sanderson 1961; Cohn 1983).  As in the $\eta=3$ case, the jet with
longitudinal B-field also develops a non-axisymmetric helical (kink)
mode close to the head of the jet which causes some beam twisting.

As expected, after propagating about the same distance, the pinches
that develop in the larger $\eta$ case are less numerous, since the
total amount of confining shocked material that deposits into the
cocoon is comparatively smaller (see Figs. 5 and 9 for a comparison).
Nonetheless, the density contrasts, $n_{is}/n_j$, attained in the pinch
regions of the MHD jet with longitudinal field, ML10r ($n_{is}/n_j
\approx 3 - 4.5$, or $n_{is}/n_{sh} \sim 0.05 - 0.07$) are
approximately the same as those in the smaller $\eta$ jet (ML3r). The
situation is a little more complex for the MHD jets with helical
field.  Over the whole jet evolution, some few pinches are found to be
stronger in the larger $\eta$ jet (MH10r) (Fig. 8), but in general the
density contrasts are approximately the same in both cases ($n_{is}/n_j
\lesssim 5$, or $n_{is}/n_{sh} \lesssim 0.13$).

\subsection{Adiabatic jets}

Fig. 10 depicts the density in the midplane section of the head of
three  supermagnetosonic, $adiabatic$ jets with $\eta=3$: a purely HD
jet (HD3a, top), an MHD jet with initial longitudinal ${\bf B}$-field,
(ML3a, middle), and an MHD jet with initial helical ${\bf B}$-field
(MH3a, bottom).  Figures 11 and 12 show the time evolution of the
corresponding velocity and magnetic field distributions. The  initial
conditions are the same as in Fig. 2 (see Table 1).  Previous numerical
analyses comparing pure HD jets with and without radiative cooling
(e.g., BFK, GB93) have shown that the presence of radiative cooling
tends to reduce the strength and number of internal shocks excited by
K-H instability. Consistently, the HD3a jet in Figs. 10-12 reveals the
appearance of a pinching zone (at $x\simeq 1R_j$) which is absent in
its radiatively cooling counterpart (HD3r, Fig. 2; see also Figs. 1 and
2 of Paper I). Besides, the beam constriction which appears close to
the head is larger in the adiabatic jet (HD3a). These results are also
compatible with the predictions of the linear stability theory (Hardee
\& Stone 1997), when applied in the context of the cooling function
employed in this work (see \S 1 ).  Similarly, the pinches which
develop in the MHD adiabatic jets (Figs. 10-12) are generally more
intense and numerous than in the radiatively cooling counterparts of
Fig. 2.  Their larger strength can be testified by direct comparison of
Figs. 13 and 5 which show the transverse profiles of some pinches along
the jets.  In particular, for the ML3a jet, we find a chain of 8
evolving pinches, at $t/t_d=1.65$, against 5 in the radiatively cooling
jet (ML3r), and the corresponding densities $n_{is}/n_j$, are about 2
to 3 times greater than in the ML3r jet.  The helical adiabatic jet
MH3a displays pinches that are 40\% denser than  those found in the
radiatively cooling jet (MH3r) at $t/t_d=1.65$.

We find similar results when comparing radiatively cooling and
adiabatic jets with  larger $\eta$. Figure 14 shows the midplane density
(left) and velocity field distribution (right) for {\it adiabatic} jets
with  $\eta=10$ after they have propagated $\simeq 33 R_j$, with the
same initial conditions as the cooling jets of figures 6, 7, and 8. The
corresponding magnetic field distribution is presented in Fig. 15.
As before, the $\eta=10$ adiabatic jet with longitudinal ${\bf
B}$-field (ML10a) develops stronger pinches than its cooling
counterpart (ML10r model) by a factor $\approx 3$. The $\eta =10$,
adiabatic helical jet (MH10a), on the other hand, has pinches with
densities of the same order of magnitude as those in the cooling jet
(MH10r).  Also, the pinches are found to be stronger in the adiabatic
jets with smaller $\eta$, in both magnetic field configurations.

\section{Conclusions and Discussion}

We have investigated here the effects of magnetic fields on the
structure of evolving overdense, radiatively cooling, supermagnetosonic
jets with the help of 3-D SPMHD simulations and compared with purely
hydrodynamical and adiabatic calculations.  Two initial magnetic field
configurations (with magnitude in approximate equipartition with the
gas) have been examined: a longitudinal, and a helical field permeating
both the jet and the ambient medium.  Calculated  for a set of
parameters which are particularly appropriate to protostellar jets
(with density ratios between the jet and the ambient medium $\eta
\approx 3-10$, and ambient Mach number $M_a \approx 24$), our results
indicate that magnetic fields have important effects on the dynamics of
radiatively cooling jets.  Both magnetic field geometries are able to
improve jet collimation relative to the pure hydrodynamical (HD) jets,
but this effect is larger in the helical field case.

As we have stressed in Paper I, the cold dense shell that develops at
the jet head due to the cooling of the shock-heated jet material in the
HD cases, also appears in the MHD jets.  Likewise, it  becomes
Rayleigh-Taylor (R-T) unstable and breaks into clumps which are more
visible in the smaller $\eta$ jets where the developed shell is
thicker.  Also as in the HD case, the shell of the MHD jets undergoes
density variations with time which are caused by global thermal
instabilities of the radiative shocks. However the amplification and
re-orientation of the nonparallel components of the magnetic fields by
the shocks at the head, particularly in the helical field geometry,
reduces the postshock compressibility and increases the postshock
cooling length. This tends to stabilize the shell against both the R-T
and the thermal instabilities. As a consequence, the clumps that are
observed to develop by fragmentation of the shell in the HD jets are
depleted in the helical field geometry. The jet immersed in the
longitudinal field, on the other hand, still retains the clumps
although they have their densities decreased relative to the the HD
counterparts.  The fact that the clumpy shell structure resembles the
knotty pattern of the Herbig-Haro objects which are commonly observed
at the  head of protostellar jets (e.g., Herbig \& Jones 1981; Brugel
et al.  1985; Reipurth 1989; Heathcote et al. 1996) suggests that a
longitudinal magnetic field geometry would be  more likely in the outer
regions of these jets than a helical field geometry (see also Paper
I).

Over the computed time and length scales, internal oblique shocks along
the beam are not found to develop in the  HD systems examined here. On
the other hand, in their MHD counterparts with both magnetic field
configurations, the confining total pressure of the cocoon excites (the
fastest growing) low-amplitude MHD K-H reflection pinching modes which
drive a chain of approximately equally spaced internal shocks along the
beam, but these shocks are found to be slightly stronger in the helical
field case (by a factor $\approx$ 20\%).  Also, as expected, the
internal shocks tend to appear in a larger number in the smaller $\eta$
jets (due to the larger amount of confining shocked material which is
deposited in the cocoon), although their densities are approximately of
the same magnitude as those in the larger $\eta$ jets.  A
non-axisymmetric helical mode is also excited close to the head of the
radiatively cooling MHD  jets with longitudinal field, which causes
some beam wiggling.

The number and strength of the internal shocks excited in the MHD
adiabatic jets is larger than in the radiatively cooling counterparts
(by a factor $\approx 2$ in both number and strength).  This result
is compatible with the linear stability theory (Hardee \& Stone 1997)
when applied in the context of the cooling function employed in the
present work, and is also compatible with previous numerical work of HD
jets which has shown that the presence of radiative cooling tends to
reduce the strength and number of internal shocks along the jet (e.g.,
BFK; GB93).  Also, the pinches are found to be more numerous in the
adiabatic jets with smaller $\eta$, in both magnetic field
configurations.

The internal shocks are found to propagate downstream with velocities
close to that of the jet head ($v_{bs} \approx 250$ km s$^{-1}$).  The
mean distance between them ($\approx 2 - 4 R_j$) is in agreement with
the observed knots in the jets. However, the weakness of the shocks in
the radiatively cooling jets ($n_{is}/n_j \approx 3 - 5$) makes it
doubtful that they could produce by themselves the bright knots
observed in protostellar jets. Probably, other mechanisms, like
intermittency in the jet injection velocity, play a more relevant role
in knot formation in those jets (see e.g., Raga et al. 1990; de Gouveia
Dal Pino and Benz 1994; Stone and Norman 1993a).

We should note that in the recent numerical study of magnetized
radiatively cooling jets by Frank et al. (1998), they have found that
toroidal fields (also in approximate equipartition with the gas) may
excite  the development of strong pinches  along the beam.  This
apparent contradiction between their analysis and ours is possibly due
to differences in the assumed initial conditions.  Frank et al. have
considered a slower and much lighter jet (with a jet Mach number
$M_{j}=v_j/c_{s_j} \approx 10$ and $\eta=1.5$) than the cases examined
here ($M_{j}=v_j/c_{s_j} \approx 42 - 76$ and $\eta=3 - 10$).  Thus,
consistently with our results above, their smaller $\eta$ jet should
be  expected to produce a larger amount of more intense pinches. This
result  has also been confirmed by numerical simulations (not presented
here) that we have performed of $\eta=1$ jets, which have produced a
larger amount of pinches with slightly larger densities than the larger
$\eta$ jets studied above.

Finally, we should make some remarks on the late evolution of the
radiatively cooling magnetized jets. In the magnetic field maps of
Figs. 2, and 8 above, we have detected the development of (sometimes
strong)  magnetic field reversals at the contact discontinuity between
the jet and the cocoon with intensities up to 5 times their initial
magnitude. As stressed in \S 3, field reversals occur in both
investigated magnetic field configurations, because the field lines are
amplified by compression in the nonparallel shocks at the jet head, and
are enforced to flow backward with the shocked plasma  into the
cocoon.  In this process, the lines  are reoriented and sometimes have
their polarization reversed.  Beyond the integrated length and time
scales depicted in the figures above, however, the increasing strength
of the reversed fields due to shear at the contact discontinuity (see
e.g., eq. 2c), may lead to the development of strong pinching regions
which ultimately may cause jet disruption, particularly in the cases
with longitudinal field. As an example, Fig. 16 depicts the time
evolution of the velocity (left) and magnetic field (right)
distributions of a disruption zone which occured in the late evolution
of the ML3r jet of Fig. 2.  We can clearly distinguish two regions with
reversed {\bf B}-fields whose strength increases with time, which are
correlated with developing pinches. The inner constriction becomes so
strong that it finally causes the disruption of the beam.  Although
shear and compression are expected to enhance {\bf B}, this
amplification is possibly partially due to numerical effects.  In fact,
along a  contact discontinuity with such a magnetic field topology with
oppositely directed  field lines pressed together, magnetic diffusion
and reconnection may have an important role and lead to intense
magnetic energy release. Of course, under the ideal-MHD approach
considered here, these dissipation effects of the magnetic field have
not been appropriately considered thus leading to possibly anomalous
amplification of the reversed components in the late stages of the
evolution of some of the jets.  Further, we should note also that no
jet disruption was detected in the majority of the adiabatic cases.
This fact is consistent with recent numerical studies of reconnection
processes in 2-D current sheets (Oreshina \& Somov 1998), which have
shown that reconnection rates are smaller  in adiabatic than in
radiatively cooling plasmas.  Although the transposition of these
results to the more complex flow geometry we have investigated  here is
not straightforward, they seem to suggest that the appropriate
consideration of magnetic field dissipation in our models will possibly
decrease and  even suppress the disruptive effects of the magnetic
fields found in some of the radiatively cooling cases examined here in
their late evolution.  The transformation of magnetic energy into
thermal energy of the gas will probably have important effects on the
structure and emission mechanisms of the beam (see e.g., Malagoli, Bodo
\& Rosner 1996; Min 1997a, b; Jones et al. 1997) and also in the
process of turbulent mixing  of the jet and cocoon material.  Such
finite magnetic resistivity effects will be addressed in a forthcoming
paper. Besides, the  potential signatures that  magnetic fields may
leave on the morphology of  radiative cooling jets, especially behind
the shocks, provide important constraints that can be used in future
observational tests to distinguish  among different candidate
mechanisms for emission line production, jet collimation, and turbulent
entrainment at the contact discontinuity between the jet and the
cocoon.

\acknowledgements
We are indebted to an anonymous referee for his fruitful comments.
A.H.C. would like to thank the Brazilian agency FAPESP, that fully
supports his work under a Ph.D. Fellowship program.  E.M.G.D.P.
aknowledges the Brazilian agencies FAPESP and CNPq for partial support.
We are also indebted to W. Benz, M. Herant, T. Jones and J. Monaghan
for useful advice and comments. The simulations were performed on a DEC
3000/600 AXP workstation, whose purchase was made possible by FAPESP.

\newpage

\centerline {\bf TABLES}

\begin{table}[ht]

\caption{The models and its initial physical parameters.}

\vspace{10pt}
\centering
\footnotesize
\tabcolsep 10pt
\begin{tabular}{ c c c c c c c c } \hline \hline 
Run & $M_a$ & $M_{A_j}$ & $M_{{ms}_j}$ & $\eta$ & $\beta$ & $q_{js}$ & $q_{bs}$ \\ \hline
HD3a  & 24 & ---  & 41.6 & 3  & $\infty$ & --- & --- \\
HD10a & 24 & ---  & 75.9 & 10 & $\infty$ & --- & --- \\
ML3a  & 24 & 38.1 & 28.0 & 3  & 1        & --- & --- \\
MH3a  & 24 & 38.1 & 28.0 & 3  & 1        & --- & --- \\
ML10a & 24 & 70.6 & 51.4 & 10 & 1        & --- & --- \\
MH10a & 24 & 70.6 & 51.4 & 10 & 1        & --- & --- \\
\hline
HD3r  & 24 & ---  & 41.6 & 3  & $\infty$ & 0.29 & 8.03 \\
HD10r & 24 & ---  & 75.9 & 10 & $\infty$ & 0.02 & 16.56 \\
ML3r  & 24 & 38.1 & 28.0 & 3  & 1        & 0.29 & 8.03 \\
MH3r  & 24 & 38.1 & 28.0 & 3  & 1        & 0.29 & 8.03 \\
ML10r & 24 & 70.6 & 51.4 & 10 & 1        & 0.02 & 16.56 \\
MH10r & 24 & 70.6 & 51.4 & 10 & 1        & 0.02 & 16.56 \\
\hline \hline
\end{tabular}
\tabcolsep 6pt
\end{table}

\newpage

\centerline {\bf FIGURE CAPTIONS}

\vskip 0.5truecm

\noindent {\bf Figure 1:} Longitudinal ($B_x$; left) and toroidal
($B_{{\phi}}$; right) magnetic field components (see equations 3 in the
text) as a function of the radial distance [$r=(y^2+z^2)^{1/2}$], for
$\beta=1$ at the jet axis. The coordinates are in code units (for the
magnetic fields 1 c.u. $= 21.5 \mu$ G).

\vskip 0.5truecm

\noindent {\bf Figure 2:} Midplane magnetic field distribution
evolution of the $\eta=3$ radiatively cooling jets ML3r (top) and MH3r
(bottom). The initial conditions are $\eta=n_j/n_a=3$, $n_a=200$
cm$^{-3}$, $M_a=24$, $v_j \simeq 398$ km s$^{-1}$, $\beta=8 \pi p/B^2
\simeq 1$, $q_{bs} \simeq 8$ and $q_{js} \simeq 0.3$.  The times and the
jet head positions are: $t/t_d=1.40$ and $x \simeq 25R_j$ (left);
$t/t_d=1.60$ and $x \simeq 29R_j$ (middle); $t/t_d=1.65$ and $x \simeq
30R_j$ (right).  Note the reorientation (and amplification), in both
models, of the magnetic fields that are carried with shocked jet
material into the cocoon.

\vskip 0.5truecm 

\noindent {\bf Figure 3:} Density of the shell at the jet axis as a
function of the time for the three radiatively cooling jets presented in
Fig. 2: HD3r (solid line); ML3r (dotted line) and MH3r (dashed line).

\vskip 0.5truecm

\noindent {\bf Figure 4:} Density (solid line) and pressure (dashed
line) across the flow at different positions along the flow in the head
region of the ML3r (top) and  MH3r (bottom) jets. The positions are in
units of $R_j$. The central peak corresponds to the beam region, while
the secondary peaks on both sides of the beam are the blobs which are
much denser in the ML3r model (top) than in the MH3r model (bottom).
These profiles are taken at $t/t_d=1.65$, and the density and pressure
scales can be calibrated using the marker in the top region of the
plots.

\vskip 0.5truecm

\noindent {\bf Figure 5:} (a) Density (solid line) and pressure (dashed
line) profiles across the HD3r (top), ML3r (middle), and MH3r (bottom)
jets taken in three different pinching positions along the flow (at
$t/t_d=1.65$).  All the profiles have been scaled as in Fig. 4. (b) The
corresponding axial density along the jet axis ($y=z=0$) showing the
channel of internal shocks. The position of the three pinches depicted
in (a) are labeled with arrows. Note that the origin of the coordinates
along the jet axis has been shifted from $-6R_j$ to 0.

\vskip 0.5truecm

\noindent {\bf Figure 6:} Gray-scale representation of the midplane
density of radiative cooling jets with $\eta=10$: a hydrodynamical jet
(top; model HD10r), an MHD jet with initial longitudinal magnetic field
distribution (middle; model ML10r) and an MHD jet with initial helical
magnetic field distribution (bottom; model MH10r), at a time
$t/t_d=1.65$. The initial conditions are $\eta=n_j/n_a=10$, $n_a=200$
cm$^{-3}$, $M_a=24$, $v_j \simeq 398$ km s$^{-1}$, $q_{bs} \simeq 16$
and $q_{js} \simeq 0.02$.  In the MHD cases, the initial $\beta \simeq
1$. The gray scale (from minimum to maximum) is given by black,
light-gray, white and dark-gray. The maximum density reached by the
shell at the head of the jets is $n_{sh}/n_a \simeq 750$ (top),
$n_{sh}/n_a \simeq 550$ (middle) and $n_{sh}/n_a \simeq 450$ (bottom).
(This and the remaining gray-scale plots below were built with the help
of the GREY subroutine developed by Marinho \& L\'epine 1998).

\vskip 0.5truecm

\noindent {\bf Figure 7:} Midplane velocity field distribution
evolution of the jets of Fig. 6: HD10r model (top), ML10r model
(middle) and MH10r model (bottom). The times and the jet head positions
are $t/t_d=1.40$ and $x \simeq 27R_j$ (left); $t/t_d=1.60$ and $x
\simeq 31R_j$ (middle); $t/t_d=1.65$ and $x \simeq 33R_j$ (right).

\vskip 0.5truecm

\noindent {\bf Figure 8:} The same as in Fig. 7 for the magnetic
field distribution of the ML10r model (top) and MH10r model (bottom).

\vskip 0.5truecm

\noindent {\bf Figure 9:} The same as in Fig. 5 for the $\eta=10$,
radiatively cooling jets: HD10r (top), ML10r (middle), and MH10r jets
(bottom), at $t/t_d=1.65$.

\vskip 0.5truecm

\noindent {\bf Figure 10:} Gray-scale representation of the midplane
density of adiabatic jets with $\eta=3$: a hydrodynamical jet (top;
model HD3a), an MHD jet with initial longitudinal magnetic field
distribution (middle; model ML3a)  and an MHD jet with initial helical
magnetic field distribution (bottom; model MH3a), at a time
$t/t_d=1.65$. The initial conditions are the same as in Fig. 6.  The
gray scale (from minimum to maximum) is given by black, light-gray,
white and dark-gray. The maximum density reached at the head of the
jets is $n_{sh}/n_a \simeq 37$ (top), $n_{sh}/n_a \simeq 39$ (middle)
and $n_{sh}/n_a \simeq 42$ (bottom).

\vskip 0.5truecm

\noindent {\bf Figure 11:}  Midplane velocity field distribution
evolution of the jets of Fig. 10: HD3a model (top), ML3a model
(middle), and MH3a model (bottom). The times and the jet head positions
are $t/t_d=1.40$ and $x \simeq 27R_j$ (left); $t/t_d=1.60$ and $x
\simeq 31R_j$ (middle); $t/t_d=1.65$ and $x \simeq 33R_j$ (right).

\vskip 0.5truecm

\noindent {\bf Figure 12:} Midplane magnetic field distribution of
the adiabatic MHD models presented in Fig. 11: ML3a (top) and MH3a
(bottom).

\vskip 0.5truecm

\noindent {\bf Figure 13:} (a) Density (solid line) and pressure
(dashed line) profiles across the HD3a (top), ML3a (middle), and MH3r
(bottom) jets taken in three different pinching positions along the
flow (at $t/t_d=1.65$). The profiles have been scaled as in Figs. 4 and
5. (b) The corresponding axial density along the jet axis ($y=z=0$),
showing the channel of internal shocks. The position of the three
pinches depicted in (a) are labeled with arrows.

\vskip 0.5truecm

\noindent {\bf Figure 14:} Gray-scale representation of the midplane
density (left) and velocity field distributions (right) of adiabatic
jets with $\eta=10$: a hydrodynamical jet (top; model HD10a), an MHD
jet with initial longitudinal magnetic field distribution (middle;
model ML10a), and an MHD jet with initial helical magnetic field
distribution (bottom; model MH10a), at $t/t_d=1.65$. The initial
conditions are the same as in Fig. 6.  The gray scale (from minimum to
maximum) is given by black, light-gray, white and dark-gray. The
maximum density reached by the shell at the head of the jets is
$n_{sh}/n_a \simeq 182$ (top), $n_{sh}/n_a \simeq 170$ (middle) and
$n_{sh}/n_a \simeq 140$ (bottom).

\vskip 0.5truecm

\noindent {\bf Figure 15:} Midplane magnetic field distribution for the
jets of Fig. 14: ML10a (top), and MH10a (bottom) at $t/t_d=1.65$.

\vskip 0.5truecm

\noindent {\bf Figure 16:} Time evolution of the velocity field (left),
and the magnetic field distribution (right) of the ML3r disrupting
region. The times shown, from top to bottom, are: $t/t_d=1.40$, 1.45,
1.49, 1.52 and 1.55.  See the text for discussion.

\end{document}